\documentclass[aps,pre,twocolumn,superscriptaddress,showpacs]{revtex4-2}

\usepackage{dcolumn}
\usepackage{bm}

\usepackage{etoolbox}

\usepackage{amsmath,amstext,amsfonts,amssymb}
\usepackage{babel}[english]
\usepackage[T1]{fontenc}
\usepackage[utf8]{inputenc}
\usepackage{graphicx}
\usepackage{upgreek}
\usepackage{gensymb}
\usepackage{color}

\usepackage{siunitx}
\DeclareSIUnit\solarmass{\ensuremath{\mathrm{M}_\odot}}
\DeclareSIUnit\year{\ensuremath{a}}
\newcommand*{\citen}[1]{%
  \begingroup
    \romannumeral-`\x 
    \setcitestyle{numbers}%
    \cite{#1}%
  \endgroup
}

\begin{document}
\title{Platform for Probing Radiation Transport Properties of Hydrogen at
Conditions Found in the Deep Interiors of Red Dwarfs}

\date{\today}

\author{J.~L\"utgert}
\email{julian.luetgert@uni-rostock.de}
\affiliation{Institut f\"ur Physik, Universit\"at Rostock, Albert-Einstein-Str.
23, 18059 Rostock, Germany}
\affiliation{Helmholtz-Zentrum Dresden-Rossendorf, Bautzner Landstrasse 400,
01328 Dresden, Germany}
\affiliation{Institute of Nuclear and Particle Physics, Technische
Universit\"at Dresden, 01069 Dresden, Germany}

\author{M.~Bethkenhagen}
\affiliation{\'Ecole Normale Sup\'erieure de Lyon, Universit\'e Lyon 1,
Laboratoire de G\'eologie de Lyon, CNRS UMR 5276, 69364 Lyon Cedex 07, France}

\author{B.~Bachmann}
\affiliation{Lawrence Livermore National Laboratory, Livermore, CA 94550, USA}

\author{L.~Divol}
\affiliation{Lawrence Livermore National Laboratory, Livermore, CA 94550, USA}

\author{D.~O.~Gericke}
\affiliation{Centre for Fusion, Space and Astrophysics, Department of Physics,
University of Warwick, Coventry CV4 7AL, United Kingdom}

\author{S.~H.~Glenzer}
\affiliation{SLAC National Accelerator Laboratory, Menlo Park, CA 94309, USA}

\author{G.~N.~Hall}
\affiliation{Lawrence Livermore National Laboratory, Livermore, CA 94550, USA}

\author{N.~Izumi}
\affiliation{Lawrence Livermore National Laboratory, Livermore, CA 94550, USA}

\author{S.~F.~Khan}
\affiliation{Lawrence Livermore National Laboratory, Livermore, CA 94550, USA}

\author{O.~L.~Landen}
\affiliation{Lawrence Livermore National Laboratory, Livermore, CA 94550, USA}

\author{S.~A.~MacLaren}
\affiliation{Lawrence Livermore National Laboratory, Livermore, CA 94550, USA}

\author{L.~Masse}
\affiliation{Lawrence Livermore National Laboratory, Livermore, CA 94550, USA}
\affiliation{CEA-DAM, DIF, F-91297 Arpajon, France}

\author{R.~Redmer}
\affiliation{Institut f\"ur Physik, Universit\"at Rostock, Albert-Einstein-Str.
23, 18059 Rostock, Germany}

\author{M.~Sch\"orner}
\affiliation{Institut f\"ur Physik, Universit\"at Rostock, Albert-Einstein-Str.
23, 18059 Rostock, Germany}

\author{M.~O.~Sch\"olmerich}
\affiliation{Lawrence Livermore National Laboratory, Livermore, CA 94550, USA}

\author{S.~Schumacher}
\affiliation{Institut f\"ur Physik, Universit\"at Rostock, Albert-Einstein-Str.
23, 18059 Rostock, Germany}

\author{N.~R.~Shaffer}
\affiliation{Laboratory for Laser Energetics, University of Rochester,
250 East River Road, Rochester, NY 14623, USA}

\author{C.~E.~Starrett}
\affiliation{Los Alamos National Laboratory, P.O. Box 1663, Los Alamos, NM
87545, USA}

\author{P.~A. Sterne}
\affiliation{Lawrence Livermore National Laboratory, Livermore, CA 94550, USA}

\author{C.~Trosseille}
\affiliation{Lawrence Livermore National Laboratory, Livermore, CA 94550, USA}

\author{T.~D\"oppner}
\affiliation{Lawrence Livermore National Laboratory, Livermore, CA 94550, USA}

\author{D.~Kraus}
\email{dominik.kraus@uni-rostock.de}
\affiliation{Institut f\"ur Physik, Universit\"at Rostock, Albert-Einstein-Str.
23, 18059 Rostock, Germany}
\affiliation{Helmholtz-Zentrum Dresden-Rossendorf, Bautzner Landstrasse 400,
01328 Dresden, Germany}

\begin{abstract}
    We describe an experimental concept at the National Ignition Facility for
    specifically tailored spherical implosions to compress hydrogen to extreme
    densities (up to $\sim 800 \times$ solid density, electron number density
    $n_e\sim \SI{4e25}{\per\cubic\centi\meter}$) at moderate temperatures
    ($T\sim \SI{200}{eV}$), i.e., to conditions, which are relevant to the
    interiors of red dwarf stars.
    The dense plasma will be probed by laser-generated x-ray radiation of
    different photon energy to determine the plasma opacity due to collisional
    (free-free) absorption and Thomson scattering.
    The obtained results will benchmark radiation transport models, which in
    the case for free-free absorption show strong deviations at conditions
    relevant to red dwarfs.
    This very first experimental test of free-free opacity models at these
    extreme states will help to constrain where inside those celestial objects
    energy transport is dominated by radiation or convection.
    Moreover, our study will inform models for other important processes in
    dense plasmas, which are based  on electron-ion collisions, e.g., stopping
    of swift ions or electron-ion temperature relaxation.
\end{abstract}
\maketitle

\section{Introduction}

Red dwarfs ($M$ dwarfs) are the lightest and coolest main sequence stars and
make up $\sim\SI{70}{\percent}$ of all stars in the Sun's
neighborhood.~\cite{Reid2005, Heath1999}
Prominent examples are our nearest neighbor Proxima Centauri
(\SI{0.12}{\solarmass}) or TRAPPIST-1 (\SI{0.089}{\solarmass}), which is only
slightly larger than Jupiter, but much more massive.
The interiors of red dwarfs mainly consist of hydrogen-helium mixtures, which
are progressively shaped by screening effects, ion-ion correlations, and
degeneracy as temperature decreases and density
increases.~\cite{Osterbrock1953}
These many-particle effects are challenging to model, in particular, for
calculations of radiation transport, which plays a major role in modeling of
sub-stellar objects and stars.
From the solar abundance problem, we know that $\sim \SI{20}{\percent}$
changes in opacity have paramount impact on our understanding of stellar
interiors.~\cite{Bailey2015,Vinyoles2017}
Whether energy can effectively be transported via radiation or, if radiation
is not sufficient, convection sets in, is a property that is particularly
influenced by stellar opacity.
In general, the physics of red dwarfs is poorly understood in comparison with
the hotter interior of the Sun, which is much closer to the ideal plasma state.

\begin{figure}
    \centering
    \includegraphics[width=1.0\linewidth]{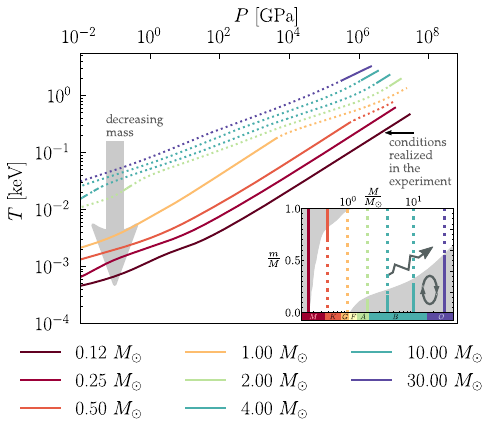}
    \caption{Pressure-temperature profiles for various celestial
    objects calculated using the MESA package.~\cite{Paxton2011, Paxton2013,
    Paxton2015, Paxton2018, Paxton2019}
    Solid lines denote convective regions while dots indicate a layer of
    radiative energy transport.
    The gray shaded area shows the conditions, which we intent to generate in
    our experiment.\newline  Inset: Mass coordinate $m/\mathrm{M}$ along
    stellar interior profiles for objects with solar composition divided into
    radiative (white, dotted lines) and convective (gray, solid curves)
    regions over total object mass $\mathrm{M}$ relative to the Sun's
    mass $\mathrm{M}_\odot$. The gray-scale dataset was published by Kippenhahn
    \textit{et al.}~\cite{Kippenhahn2012} while the colored lines show the MESA
    calculations of the main figure.}\label{fig:stars}
\end{figure}

Figure~\ref{fig:stars} shows simulated pressure-temperature profiles of stars on
the main sequence, demonstrating the extreme plasma conditions present inside
those celestial objects.
The curves were obtained using the MESA code for stellar
evolution~\cite{Paxton2011, Paxton2013, Paxton2015, Paxton2018, Paxton2019}
(see the appendix for details on these simulations).
The inset illustrates the schematic interiors of stars from the core
($m/\mathrm{M}=0$) to the photosphere ($m/\mathrm{M}=1$) divided into radiative
and convective zones for stars with a solar composition of elements.
Red dwarfs are characterized by masses between 0.075 and 0.5 solar masses so
that their typical mass-temperature ratios overall place red dwarfs in a
regime, where convection dominates the outer regions.
Depending on the size of the individual object, a more or less developed
radiative core is present.
The smallest red dwarfs ($\mathrm{M} \lesssim\SI{0.2}{\solarmass}$) are
thought to be fully convective.
In this case, the fusion reactions in the core are permanently re-fueled by
hydrogen from the outer layers.
Combined with the low fusion rates due to the relatively low core temperatures,
convection possibly allows some red dwarfs to last trillions of years until
all hydrogen fuel is exhausted.
However, even a small radiative core can strongly change this behavior and its
existence crucially depends on the effectiveness of radiation transport in
highly compressed matter.

Moreover, the internal structure of a star has a major impact on the activity
of its surface.~\cite{Babcock1961}
The boundary between a radiative core and a convective layer can lead to strong
magnetic fields and a turbulent atmosphere,~\cite{Parker1975, Reid2005,
Route2016} including radiative and plasma outbursts that may threaten life on
nearby planets.
Therefore, understanding the radiative properties of the complex plasmas within
a host star is crucial when judging the possibility of an exoplanet to host
life -- especially for red dwarfs where the habitable zone is thought
to be found relatively close to the star itself due to the low surface
temperature.~\cite{Parke_Loyd2018}

For red dwarf stars, the thermodynamic conditions at the boundary between
radiative core and convective envelope are estimated to be in a pressure
regime of few Gbars and temperatures of few million
Kelvin.~\cite{Kippenhahn2012,Reid2005}
Corresponding free electron densities are in the range of few
\SI{e25}{\per\cubic\centi\meter}, which results
in Fermi energies of similar order as the thermal energy of the free electrons.
The energy transport in this so-called warm dense matter
regime~\cite{Graziani2014} is extremely difficult to calculate, which gives
rise to significant uncertainties in modeling the energy transport inside red
dwarfs.

\section{Theory}

For stellar interiors, the radiative opacity $\kappa_\text{rad}$ is usually
divided into three contributions:~\cite{Hayashi1962}
\begin{equation}
\kappa_\text{rad}=\kappa_{bf}+\kappa_{ff}+\kappa_T,
\end{equation}
where $\kappa_{bf}$ is the opacity contribution by bound-free absorption,
$\kappa_{ff}$ denotes the free-free contribution and $\kappa_T =
Z\sigma_T/m_i$ the absorption due to Thomson scattering from free electrons,
which is solely dependent on the Thomson scattering transport cross section
$\sigma_T$,~\cite{Boercker1987, Watson1969} the average ion charge state $Z$
and the average ion mass $m_i$ of the plasma.
While Rayleigh scattering might be of interest for the atmosphere of $K$ and
$M$ class stars,~\cite{Hubeny2014} the high ionization in hotter photospheres
and deep within even the smallest stars often justifies to neglect its
contribution to $\kappa_\text{rad}$.
For the solar abundance problem, the bound-free opacity of metals is probably
most relevant, but deep in the solar radiation zone as well as for many red
dwarfs, particularly those with low metallicity, hydrogen free-free opacity,
i.e., absorption due to inverse bremsstrahlung, is the dominant absorption
mechanism of radiation.~\cite{Hayashi1962}
For these red dwarfs, the absolute values for free-free absorption determine
where convection or radiation will be the dominant energy transport mechanism.

\begin{figure}
    \centering
    \includegraphics[width=1.0\linewidth]{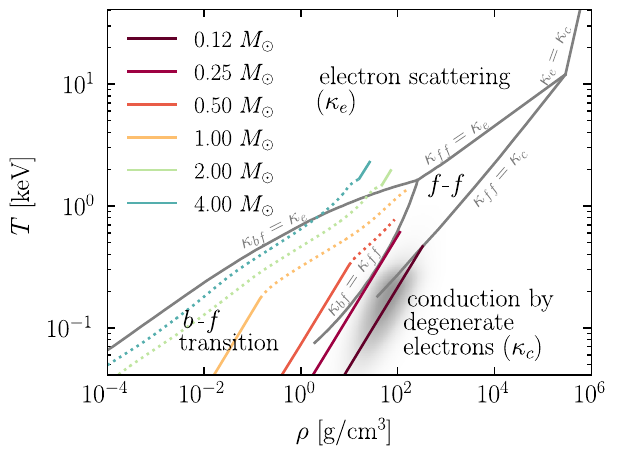}
    \caption{Dominating opacities in different regimes of the
    density-temperature diagram for a composition of elements as in
    population I stars.~\cite{Hayashi1962}
    The colored lines show densities and temperatures realized within
    main-sequence stars according to  the ``MESA'' stellar evolution
    code~\cite{Paxton2011, Paxton2013, Paxton2015, Paxton2018, Paxton2019} with
    solid lines representing convective layers.
    The proposed experiments will probe conditions similar to the interiors of
    red dwarf stars where free-free absorption is expected to dominate
    (indicated by the shaded region).}\label{fig:dominant_opacities}
\end{figure}

Figure~\ref{fig:dominant_opacities} shows a density-temperature diagram of
dominating absorption mechanisms thought to be present for the composition of
population I stars in comparison with red dwarf interiors and the Sun.
While bound-free transitions dominate at low densities and temperatures,
free-free absorption starts to outrun the bound-free opacity with the increase
in density due to increasing electron-ion collision rates as well as pressure
ionization of heavier elements.~\cite{Clayton1983}
At the highest densities, conduction by degenerate electrons becomes more
efficient than radiation transport, whereas for low densities and highest
temperatures, photon scattering from electrons (Thomson or Compton, depending
on photon energy) is most significant.

\subsection{Analytical models}

A classical treatment of the spectral absorption coefficient due to inverse
bremsstrahlung, derived from the description of electron-ion collisions in a
weakly coupled plasma environment, yields for the absorption coefficient
$\alpha_{ff}$,~\cite{Kramers1923}
\begin{equation}
  \alpha_{ff}(\nu) = \rho \kappa_{ff}(\nu) \propto \frac{Z^2 n_e n_i}{\nu^3
  \sqrt{T}} \left[1 - \exp\left(-\frac{h \nu}{k_B T}\right)\right]
  g_{ff}(\nu,T),
  \label{eqn:ffabsorb}
\end{equation}
where $\nu$ denotes the x-ray frequency, $\rho$ is the mass density, $Z$ is the
average degree of ionization, $n_e$ is the free electron number density, $n_i$
is the ion number density, $T$ is the plasma temperature, $h$ is Planck's
constant, and $k_B$ is Boltzmann's constant.
Additional corrections due to quantum and correlation effects are accounted
for in a frequency-dependent correction factor $g_{ff} (\nu,T)$, the so-called
Gaunt factor.~\cite{Gaunt1930}
For a weakly coupled ideal plasma, the Gaunt factor can be interpreted as the
logarithm of the ratio of maximum impact parameter $b_\text{max}$ and minimum
impact parameter $b_\text{min}$ in the corresponding electron-ion collision
(the so-called Coulomb logarithm~\cite{Meyer-ter-Vehn2019}):
\begin{equation}
    g_{ff}(\nu,T) = \frac{\sqrt 3}{\pi} \ln
    \left(\frac{b_\text{max}}{b_\text{min}}\right).
\end{equation}
This formalism is equivalent to the classical treatment of several other
important plasma effects that involve Coulomb collisions of electrons and
ions, e.g., stopping power of ions or electron-ion temperature equilibration in
dense plasmas.~\cite{Gericke2002}
The maximum impact parameter $b_\text{max}$ is usually given by
$\min(v_e/2\pi \nu,\lambda_s)$ where $v_e$ is the average velocity of the
electrons, $\nu$ is the x-ray frequency, and $\lambda_s$ is the screening
length due to the surrounding plasma.
On the other hand, $b_\text{min}$ can be expressed as
$\max(b_\perp, \lambda_{th}$), where $b_\perp=Ze^2/(4\pi \epsilon_0 m_e v_e^2)$
is the impact parameter for an electron being deflected perpendicular to its
direction of incidence and $\lambda_{th}$ denotes the thermal de Broglie
wavelength of the electrons.

However, for conditions relevant to the interiors of red dwarfs [e.g.,
$n_e\sim$ few $\SI{e25}{\per\cubic\centi\meter}$, $T_e
\sim$  few $ \SI{100}{eV}$ (Ref.~\citen{Osterbrock1953})], we find $v_e/2\pi\nu <
\lambda_{th}$, i.e., a negative Coulomb logarithm for x-ray frequencies larger
than the plasma frequency.
Thus, this simple classical treatment assuming a weakly coupled plasma is not
appropriate for such conditions.
Indeed, more sophisticated approaches have been developed to accommodate these
conditions.~\cite{Grinenko2009, Wierling2001}

\subsection{Average Atom and Density Functional Theory calculations}

\begin{figure}
    \centering
    \includegraphics[width=1.0\linewidth]{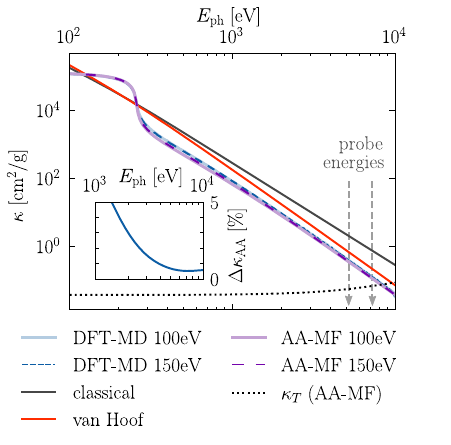}
    \caption{Opacity for a (\SI{25}{\percent}/\SI{75}{\percent}) HT mixture at
        an electron density $n_e=\SI{5e25}{\per\cubic\centi\meter}$ and
        a temperature of $T=\SI{100}{\electronvolt}$ (solid and dotted lines)
        or $T=\SI{150}{\electronvolt}$ (dashed), respectively, according to
        different models.
        The blue and purple lines depict DFT-MD and AA-MF calculations.
        The other solid curves show the analytical model for free-free
        extinction [Eq.~(\ref{eqn:ffabsorb})] with a Gaunt factor equal to
        unity (black line) or values calculated by van Hoof \textit{et
        al.}~\cite{VanHoof2014, VanHoof2015} (red).
        For comparison, the opacity due to Thomson scattering for a temperature
        of $T=\SI{100}{\electronvolt}$ is shown as the dotted black line.
        The AA-MF calculation at \SI{150}{\electronvolt} and the inset
        depicting the relative difference between the two results
        [$\Delta\kappa_\text{AA} = 2
        (\kappa_\text{AA}^{\SI{150}{\electronvolt}} -
        \kappa_\text{AA}^{\SI{100}{\electronvolt}})
        /(\kappa_\text{AA}^{\SI{150}{\electronvolt}} +
        \kappa_\text{AA}^{\SI{100}{\electronvolt}})$] show that the
        temperature influence is very small for photon energies above
        \SI{3}{\kilo\electronvolt} due to the degenerate
conditions.}\label{fig:opacities_calc} \end{figure}

Figure~\ref{fig:opacities_calc} shows Average Atom calculations with a Mean Force
potential (AA-MF)~\cite{Starrett2017} and state-of-the-art Density Functional
Theory Molecular Dynamics (DFT-MD) simulations~\cite{Bethkenhagen2020} compared
to the analytical model for the free-free opacity with constant Gaunt factor
and calculations by van Hoof \textit{et al.}~\cite{VanHoof2014,VanHoof2015}
The first two simulation methods have previously been applied successfully for
calculating the equation-of-state (EOS) and transport properties of dense
plasmas.~\cite{Becker2014,Becker2018,Holst2011,Starrett2017}
The DFT-MD simulations were performed with up to 256 hydrogen atoms using the
program package VASP.~\cite{Kresse1993,Kresse1994,Kresse1996b,Kresse1996}
Our considered density range spans
\SI{20}-\SI{150}{\gram\per\cubic\centi\meter} at \num{100},
\num{150}, and \SI{200}{\electronvolt}.
The simulations use the Baldereschi mean value point~\cite{Baldereschi1973} and
the Coulomb potential with an energy cutoff of \SI{10000}{\electronvolt}. Each
DFT-MD point was run for \num{20000} time steps with a time step size between
\SI{3}{\atto\second} (attoseconds) and \SI{8}{\atto\second} depending on the
thermodynamic conditions.
The temperature was controlled with a Nos\'{e}-Hoover
thermostat.~\cite{Nose1984}
Subsequently, \numrange{10}{20} snapshots were selected from each trajectory to calculate
the electrical conductivity and opacity applying the Kubo-Greenwood
formalism~\cite{Kubo1957, Greenwood1958} and the Kramers-Kronig relation.
For details on the AA-MF calculations (which were performed for identical
temperatures and pressures), we refer to previous
publications.~\cite{Starrett2013,Starrett2015,Starrett2016,Starrett2017,Shaffer2017}

While the DFT-MD simulation naturally includes many-body effects in the
description of wavefunctions and the density of states due to the multiple ions
included in the simulation, the AA-MF approach, which is strictly
speaking also DFT-based, simplifies the calculation by exclusively relying on
the atom-in-jellium model.
Either of the two formulations calculates opacity from the real part of the
electrical conductivity.
Both approaches agree on the quantity of extinction remarkably well, supporting
each other.
This result is particularly noteworthy as the similarity of the AA-MF
calculation with DFT-MD -- for the specific case of hydrogen -- is highly
desired: While DFT-MD is generally more accurate, AA-MF is computationally
significantly less expensive and should be favored if benchmarks can show good
agreement between the two methods.
At the same time, the consistency of the computed opacity illustrates
impressively that extreme states of matter can be treated by DFT-MD
nowadays with the increase in computational power and, hence, number of energy
bands included in the calculation.

Both methods show a discrepancy to the calculation of the free-free opacity
from the semi-classical formula [Eq.~(\ref{eqn:ffabsorb})], as it can be seen
in Fig.~\ref{fig:opacities_calc}. The simplest approach of setting the Gaunt
factor to unity reproduces the classical result of Kramers.~\cite{Kramers1923}
Introducing quantum-mechanical corrections, van Hoof \textit{et
al.}~\cite{VanHoof2014, VanHoof2015} provide easily applicable, tabulated
values for $g_{ff}$ by following the seminal work of Karzas and
Latter.~\cite{Karzas1961}
However, this calculation is requiring more assumptions than the AA-MF or the
DFT-MD model, e.g., the velocity distribution of the electrons (with is assumed
to be Maxwellian) in order to calculate thermally averaged free-free
Gaunt factors and from these opacities.~\cite{VanHoof2014, VanHoof2015} Other
authors perform similar calculations but integrate over the Fermi distribution
of a degenerate electron gas.~\cite{Meyer-ter-Vehn2019}

In fact, for dense plasma conditions comparable to the interiors of main
sequence stars, even advanced calculations of the Gaunt factor vary by more
than \SI{50}{\percent} for frequencies larger than the plasma
frequency.~\cite{Wierling2001}
In particular, the specific treatment of dynamic screening, strong collisions,
and re-normalization due to higher moments can make a significant difference in
comparison to widely-used Born approximation treatments of the Gaunt
factor.~\cite{Wierling2001}
Deviations are particularly significant in the photon energy regime from
$\sim\SI{500}{\electronvolt}$ to few \unit{\kilo\electronvolt}, which is the
dominant contribution when calculating the Rosseland mean opacity for the Sun
and smaller stars.
Indeed, varying the opacity by \SI{50}{\percent} can change the radii of the
boundaries between convection and radiation zone by up to \SI{10}{\percent},
which, given the underlying density and temperature gradients, would
significantly impact our general understanding of
stars.~\cite{Vinyoles2017,Serenelli2011}

\section{Experimental Concept}
Using the largest laser system in the world, namely, the National Ignition
Facility (NIF) at Lawrence Livermore National Laboratory,~\cite{Moses2009} it
is now possible to create and probe matter states relevant to stellar interiors
in the laboratory.~\cite{Kritcher2020,Kraus2016b}
To address the questions raised above, we have developed a concept to
leverage NIF's unique capabilities to create relevant conditions and obtain a
very first test of free-free opacity models in this very important plasma
regime via x-ray absorption measurements of highly compressed hydrogen during
the stagnation phase of layered capsule implosions.
In this way, not only the various existing models and resulting tables for the
Gaunt factor will be tested, but also modern DFT-MD and AA-MF simulations,
which provide the absorption coefficient, can be benchmarked.
Finally, due to the equivalent physics involved (electron-ion collisions in
dense plasma environments~\cite{Grinenko2009}), our results on free-free
absorption will inform models for swift ion stopping in warm dense matter as
well as corresponding electron-ion equilibration times.

\subsection{Experimental setup}
\begin{figure}
    \centering
    \includegraphics[width=\linewidth]{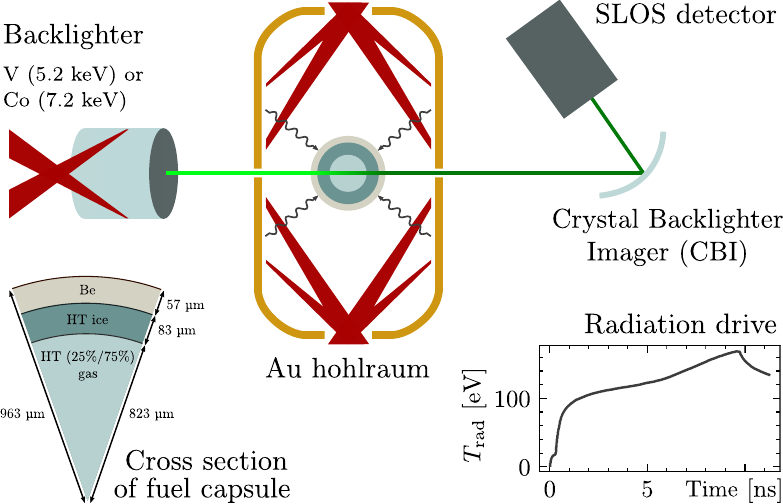}
    \caption{Schematic of the experimental setup.
        Using NIF's laser beams, a nearly Planckian x-ray bath (see bottom right) is created by
        heating a gold \textit{Hohlraum} with a fuel capsule  at its center. Upon stagnation, the solid hydrogen layer is compressed to mass densities larger than \SI{100}{\gram\per\cubic\centi\meter}. Two \textit{Hohlraum} windows
        allow for measuring the transmission of high density hydrogen using x rays created in a stagnating plasma inside the backlighter tube.
        Fielding NIF's Crystal Backlighter Imager \cite{Hall2019} and the single
        line-of-sight (SLOS) detector \cite{Engelhorn2018} enables us to
        acquire narrow bandwidth high-resolution radiography images of the implosion.}\label{fig:setup}
\end{figure}

A sketch of the experimental setup is shown in Fig.~\ref{fig:setup}. We will
use 184 out of the 192 NIF laser beams to heat a gold \textit{Hohlraum}
creating a quasi-thermal radiation field that implodes a layered fuel capsule
at the center of the \textit{Hohlraum}. The capsule is comprised of a
\SI{57}{\micro\meter} thick beryllium ablator shell, containing a
\SI{83}{\micro\meter} thick layer of cryogenic solid hydrogen.

The temporally shaped radiation field created by the laser drive will ablate
the Be shell and, hence, accelerate the payload inward. Upon stagnation, a high
density hydrogen layer with $\rho > \SI{100} {\gram\per\cubic\centi\meter}$ is
formed while most of the Be ablator has been ablated.

The implosion design is derived from inertial confinement fusion (ICF)
implosions at the NIF~\cite{ZylstraPoP2019} and applies a well-tested model that matched a variety of spherical DT implosion experiments in NIF's ICF program.~\cite{KritcherPoP2018,ZylstraPoP2019}
In contrast to ICF implosions aiming for high neutron yield,\cite{ZylstraN2022} the peak radiation temperature of
our \textit{Hohlraum} drive ($T_\text{rad}=\SI{170}{\electronvolt}$) is significantly reduced,
deliberately slowing down the implosion to
$\sim\SI{200}{\kilo\meter\per\second}$ with the goal of creating extreme
densities ($\sim\SI{150}{\gram\per\cubic\centi\meter}$) at moderate
temperatures ($\sim\SI{200}{\electronvolt}$) while reducing x-ray and
neutron-related background signals near stagnation that would affect the
radiography measurement.
These conditions are directly relevant to the interiors of red dwarfs.

The dense plasma states will be probed by x-ray line emission, which
is created by the eight remaining NIF beams that heat a backlighter tube. We will perform 2D-imaging radiography of the implosion and stagnation phase
using two different photon energies by varying the backlighter tube material.
Vanadium will result in \SI{5.2}{\kilo\electronvolt} line emission from
1s2p$\rightarrow$1s$^2$ (He-$\upalpha$) transitions of helium-like ions.
In the same way, helium-like cobalt ions will produce
\SI{7.2}{\kilo\electronvolt} photons.

NIF's Crystal Backlighter Imager (CBI) system~\cite{Hall2019} will allow for high-resolution, nearly monochromatic radiography images of the imploding and stagnating sphere.
A gated single-line-of-sight (SLOS) detector~\cite{Engelhorn2018} will provide four images per implosion with a time delay of $\sim\SI{100}{\pico\second}$
between consecutive images.  From the radiography images, we will infer radial profiles of the opacity through Abel inversion~\cite{Hanson1989,Howard2016} and forward fitting methods.

\subsection{Radiation hydrodynamic simulations}
    \begin{figure}
    \centering
    \includegraphics[width=\linewidth]{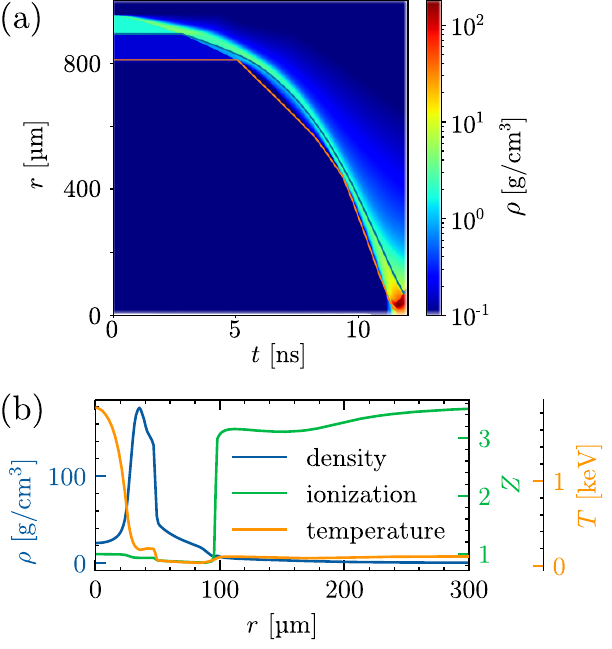}
    \caption{(a) 1D radiation-hydrodynamics simulations performed using the
        HYDRA code,~\cite{Marinak2001} showing the mass density $\rho$ at a
        given radius $r$ and time $t$. Our experiment intends to probe
        around stagnation when the highest fuel compression is
        predicted.
        Density, temperature, and ionization profiles around this time
        ($\sim\SI{11.5}{\nano\second}$ after the start of the drive laser) are
        plotted in (b) and predict hydrogen mass densities in excess of
        \SI{150}{\gram\per\cubic\centi\meter} in the HT mixture with a molar
        mass of \SI{2.5}{\gram\per\mole}.
    }\label{fig:design}
\end{figure}

The reduced implosion velocity minimizes the hot spot x-ray emission at
stagnation to improve the signal-to-noise ratio of the physics measurements.
For the same reason, we use a (\SI{25}{\percent}/\SI{75}{\percent}) HT mixture
(which is hydrodynamically equivalent to \SI{50}{\percent}/\SI{50}{\percent}
DT) for the ice layer to minimize the neutron yield
and the related background signals. Tritium is required to enable the hydrogen
ice layer formation through beta layering, where self-heating from beta decay
leads to a redistribution of HT ice with time.~\cite{Glenzer2012, Martin1988,
Mruzek1988}
Figure~\ref{fig:design} shows an overview of the $\sim\SI{11}{\nano\second}$
implosion trajectory simulated with HYDRA-1D.~\cite{Marinak2001}
Compared to previous ICF experiments with Be shells, \cite{ZylstraPoP2019} the
ablator thickness is reduced to \SI{57}{\micro\meter} to decrease the remaining
ablator mass to near zero,
which will maximize the radiography contrast of the hydrogen layer.
Figure~\ref{fig:design}(b) illustrates examples of radial density, ionization,
and temperature profiles predicted by hydrodynamic simulations.
The high-density portion of the profiles consists of hydrogen (HT) only.
Our simulations predict that the material is compressed to mass densities
exceeding $\SI{150}{\gram\per\cubic\centi\meter}$, which corresponds to
electron densities of $\SI{3.6e25}{\per\cubic\centi\meter}$, at a temperature
of $\sim\SI{200}{\electronvolt}$. These parameters result in Fermi energies of
$\sim\SI{400}{\electronvolt}$ and, thus, $\sim2\times$ larger than the thermal
energies.
Hence, we will probe degenerate plasma conditions as expected in the interiors
of red dwarf stars.
In addition, these conditions are expected to limit the influence of
temperature on the free-free absorption, since the $1/\sqrt{T}$ term in the
expression for the inverse bremsstrahlung [Eq.~(\ref{eqn:ffabsorb})] can
be replaced approximately by $1/\sqrt{T_F}$, where $T_F$ denotes the Fermi
temperature.
This behavior is supported by the results from DFT-MD and the Average Atom
Compared to previous ICF experiments with Be shells,~\cite{ZylstraPoP2019} the ablator
thickness is reduced to \SI{57}{\micro\meter} to decrease the remaining ablator
mass to near zero,
model, which indicates that the opacity is more sensitive to the Fermi
temperature than the electronic temperature under dense degenerate plasma
conditions (see Fig.~\ref{fig:opacities_calc} and inset).
The weak dependence on $T$ is highly beneficial for the analysis and
interpretation of our data, as we can determine opacity and from there infer
density, while a direct measurement of temperature would require additional
diagnostics -- for example x-ray Thomson scattering.~\cite{Glenzer2009}

\begin{figure}
    \centering
    \includegraphics[width=\linewidth]{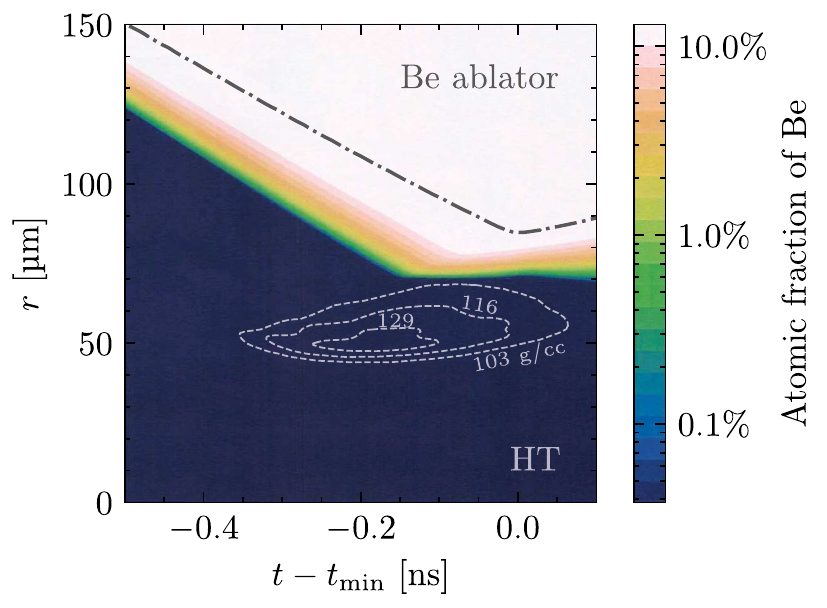}
    \caption{Beryllium content of the capsule close to stagnation as a function
    of time $t$ and radius $r$. The dash-dotted dark line shows the interface of
    ablator and fuel for a simulation without mixing. The dashed lines depict
    contours of constant density in the highly compressed HT. For the times we
    intend to probe (around \SI{-0.4}{\nano\second} to \SI{-0.2}{\nano\second}), the Be is
    not predicted to contaminate this dense part of the fuel.
    The values on the abscissa are given relative to the time where the
    rebounding shock wave coincides with the fuel-ablator interface.
}\label{fig:mixing}
\end{figure}

Reducing the ablator shell thickness and decreasing the remaining mass at
stagnation might come with increased risk of hydrodynamic instability at the
ablator-ice interface and ablator material mixing into the compressed ice layer.
%
%
For a more quantitative estimate of mixing, we have performed 1D capsule-only
simulations using a  buoyancy-drag mix model,~\cite{Dimonte2000} which was
successful in explaining experimental performance observations of previous
layered Be implosions.~\cite{ZylstraPoP2019} In addition, more recently the
buoyancy-drag mix model has been calibrated in a focused series of experiments
using a thin ice layer of varying thickness and detecting neutronic signatures of
deuterated plastic, originally located near the inside of a plastic shell,
mixing through the ice layer into the hot spot.~\cite{Bachmann2022}
Figure~\ref{fig:mixing} shows the results for our significantly slower implosion
design in terms of atomic Be fraction as function of radius and time. The
demarcation line between the Be ablator and the HT ice is indicated as a
dot-dashed line. We have labeled the time of minimum radius of this interface
by $t_\text{min}$, which coincides with the time when the rebounding shock wave,
which leads to the formation of the high-compression HT ice, passes this
interface.
Our simulations clearly show that Be does not mix into the high-compression HT
ice layer. The radiography measurements are aimed to record transmission images
between 400 and \SI{200}{\pico\second} before $t_\text{min}$, which, hence, is
not affected by Be mixing into the high compression HT ice layer.

\section{Simulated radiography signal}

\begin{figure}
    \centering
    \includegraphics[width=\linewidth]{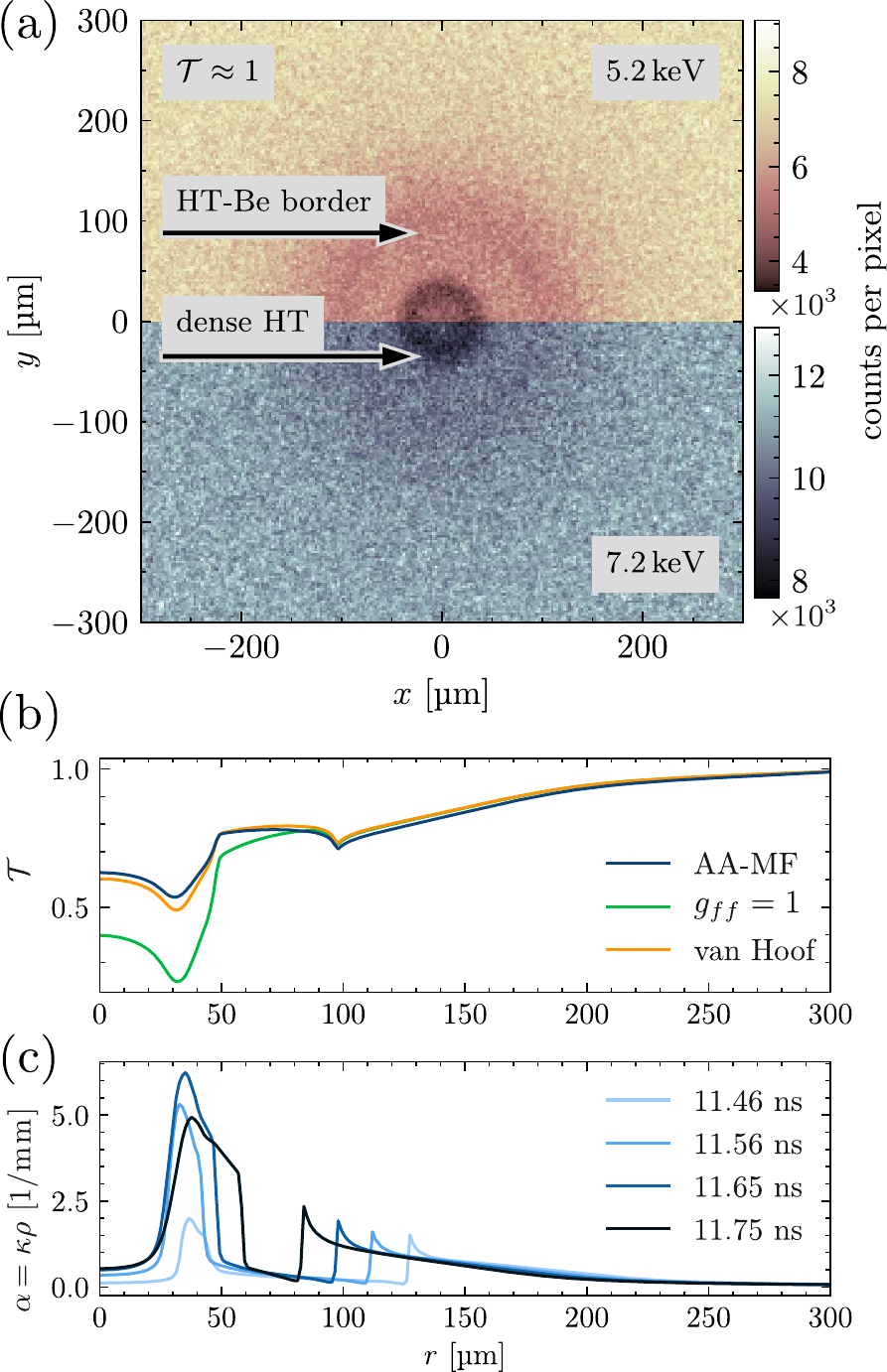}
    \caption[(a) Simulated detector image including free-free (AA-MF calculations)
    and bound-free~\cite{Verner1995} absorption as well as
    Thomson scattering~\cite{Boercker1987}
    for \SI{5.2}{\kilo\electronvolt} (top) and \SI{7.2}{\kilo\electronvolt}
    (bottom) backlighter energy~
    The colorbar indicates the expected number of photons per pixel.
    Albeit the limited temporal and spatial resolution of the detector, the
    interface between the HT mixture and the beryllium ablator is predicted to
    be visible at least for the lower energy.
    (c) Absorption coefficient profiles and (b) integrated total capsule
    transmission $\mathcal{T}$ for a \SI{5.2}{\kilo\electronvolt} backlighter
    close to the time where the highest density in the hydrogen is reached.
    As the lower figure shows, different models for the free-free opacity
    result in notable changes of the total transmission. All presented plots
    assume a perfectly symmetrical implosion, no mixing between fuel and ablator
    and a beryllium layer without impurities.
    ]{(a) Simulated detector image including free-free (AA-MF calculations)
    and bound-free~\cite{Verner1995} absorption as well as
    Thomson scattering~\cite{Boercker1987}
    for \SI{5.2}{\kilo\electronvolt} (top) and \SI{7.2}{\kilo\electronvolt}
    (bottom) backlighter energy.
    The colorbar indicates the expected number of photons per pixel.
    Albeit the limited temporal and spatial resolution of the detector, the
    interface between the HT mixture and the beryllium ablator is predicted to
    be visible at least for the lower energy. (c) Absorption coefficient
    profiles and (b) integrated total capsule transmission $\mathcal{T}$ for a
    \SI{5.2}{\kilo\electronvolt} backlighter close to the time where the
    highest density in the hydrogen is reached.  As the lower figure shows,
    different models for the free-free opacity result in notable changes of the
    total transmission. All presented plots assume a perfectly symmetrical
    implosion, no mixing between fuel and ablator and a beryllium layer without
    impurities.
    }\label{fig:lineouts}
\end{figure}


Synthetic radiography images (see Fig.~\ref{fig:lineouts} and the appendix for
more details) show a clear limb feature at the boundary of the central hydrogen
and the beryllium layer.
Applying mass conservation, this feature will be used as constraint for the
density of the encapsulated hydrogen at smaller radii as the total initial fuel
mass can be accurately characterized.
Toward the outer edges of the radiography field-of-view of
$\num{600}\times\SI{600}{\square\micro\meter}$, we expect the plasma
to become fully transmissive to the probe radiation, which will be used to
normalize the opacities measured at smaller radii and obtain absolute values
of the radial absorption coefficient.

In the investigated density and temperature regime, opacity due to Thomson
scattering is expected to reach similar values as free-free absorption.
As absorption due to Thomson scattering scales linearly with $n_e$ and
free-free opacity with $n_e^2$, this effect can become particularly
significant in the lower density regions of the imploded capsule.
To disentangle both mechanisms, we will perform the absorption measurements at
two photon energies (\SI{5.2}{\kilo\electronvolt} and
\SI{7.2}{\kilo\electronvolt} as described above).
While disagreeing in the actual magnitude, all but the classical approach
($g_{ff} = 1$) presented in Fig.~\ref{fig:opacities_calc} find the same
proportionality of the free-free opacity with $\sim 1/{(h\nu)}^{(7/2)}$ for
high photon energies $h\nu$.
Classically, the Thomson cross-section is in fist order independent of
probing frequency and temperature, which would allow us to separate the two
contributions to the signal. While this assumption might give a reasonable
first estimate, our AA-MF simulations (see Fig.~\ref{fig:opacities_calc})
indicate in accordance with previous calculations by
Boercker~\cite{Boercker1987} that this description as a constant is not
applicable at the extreme conditions we intend to probe. However, models of the
Thomson scattering have to provide only the ratio between the cross-section at
the two backlighter energies to enable us to differentiate the former from
inverse bremsstrahlung.
Regardless of the model applied, the Thomson scattering's contribution to the
overall opacity might also be used as an additional density constraint next to
measuring the ablator-hydrogen interface and applying mass conservation.

Further constraints on the implosion parameters will be deduced from measuring
multiple absorption images at different times during the implosion in one shot.
The stagnation phase is usually well modeled by a self-similar description of
the conservation laws,~\cite{Atzeni2004} which only allows certain shapes and
time evolutions of the density profiles.

\section{Conclusions}
In summary, we presented a concept to leverage NIF's unique capabilities to
investigate the deep interiors of red dwarf stars in the laboratory and shed
light on their internal energy transports mechanisms.
The proposed experiment has been accepted within NIF's Discovery Science
Program for upcoming shot days in 2022 and 2023.
The resulting measurement of free-free absorption and opacity will provide a
benchmark for numerical and analytical approaches, which will in turn
yield an improved description of the Gaunt factor.
Finally, interior structure models for massive hydrogen-rich astrophysical
objects, such as red dwarfs, can be revisited on the basis of the new opacity
constraint.

\section*{Acknowledgments}
M.B. was supported by the European Horizon 2020 programme within the Marie
Sk\l odowska-Curie actions (xICE grant number 894725).
J.L. and D.K. acknowledge support by the Helmholtz Association under
VH-NG-1141 and by GSI Helmholtzzentrum für Schwerionenforschung,
Darmstadt as part of the R\&D project SI-URDK2224 with the University of
Rostock.
The work of S.S. and D.K. was supported by Deutsche Forschungsgemeinschaft
(DFG – German Research Foundation) project no. 495324226.
The work of B.B., L.D., G.N.H., S.F.K., N.I., O.L.L., S.A.M, L.M., M.O.S.,
P.A.S. and T.D. was performed under the auspices of the DOE by Lawrence
Livermore National Laboratory under Contract No. DE-AC52-07NA27344.
R.R. and M.S. acknowledge support from the DFG via the
Research Unit FOR 2440.
The DFT-MD calculations were performed at the North-German
Supercomputing Alliance (HLRN) facilities and at the IT- and Media Center of
the University of Rostock.
\section*{Author declarations}
\subsection*{Conflict of interest}
The authors have no conflicts of interest.
\section*{Data availability}
The data that support the findings of this study are available
from the corresponding authors upon reasonable request.

\appendix
\section{MESA Simulations}
Profiles of temperature, density, and pressure inside stars with various masses
were calculated using the ``Modules for Experiments in Stellar Astrophysics''
code (MESA, release r15140).~\cite{Paxton2011, Paxton2013, Paxton2015,
Paxton2018, Paxton2019}
We used the default MESA equation-of-state (EOS) which is a blend of the
OPAL,~\cite{Rogers2002} SCVH,~\cite{Saumon1995} FreeEOS,~\cite{Irwin2004}
HELM,~\cite{Timmes2000} and PC~\cite{Potekhin2010} EOSes.
Radiative opacities are primarily from OPAL,~\cite{Iglesias1993,
Iglesias1996} with low-temperature data from Ferguson \textit{et al.}~\cite{Ferguson2005} and the
high-temperature, Compton-scattering dominated regime by Buchler and Yueh.~\cite{Buchler1976}
Electron conduction opacities are from Cassisi \textit{et al.}.~\cite{Cassisi2007}
Nuclear reaction rates are from JINA REACLIB~\cite{Cyburt2010} plus
additional tabulated weak reaction rates.~\cite{Fuller1985, Oda1994,
Langanke2000}
Screening is included via the prescription of Chugunov \textit{et al.}.~\cite{Chugunov2007}
Thermal neutrino loss rates are from Itoh \textit{et al.}.~\cite{Itoh1996}

A pre-main-sequence model has been calculated from initial parameters of helium mass
fraction $Y_i = 0.2744$, metallicity $Z_i = 1-X_i-Y_i = 0.01913$ (with $X_i$
being the initial mass fraction of hydrogen), mixing length parameter
$\alpha_\text{MLT} = 1.9179$, and including element diffusion, which has been
found to reproduce the solar model well.~\cite{Paxton2011}
The ratio of elements heavier than helium was taken from Grevesse and Sauval.~\cite{Grevesse1998}
The time span of the simulation was chosen so that the star spent a
considerable amount of time on the main sequence.
For stars smaller than the Sun, 10 times the age of the star when the main
sequence was entered has been chosen; masses $\mathrm{M}>\mathrm{M}_\odot$
were evolved until $t_\text{nuc}/2$ was reached, were $t_\text{nuc} =
{\left(\mathrm{M}/\mathrm{M}_\odot \right)}^{-2.9} \times \SI{e10}{\year}$ is
an approximation for the lifetime of star on the main
sequence.~\cite{Hansen2004}

The decision whether regions of a star were considered convective or radiative
was based on the Schwarzschild criterion.

\section{Comparing radiative and conductive opacities}
The relation between thermal conductivity through photon transport (radiation)
$\lambda_\text{rad}$ and the opacity $\kappa_\text{rad}$ is given by
\begin{align}
    \lambda_\text{rad} = \frac{4 a c T^3}{3 \rho \kappa_\text{rad} }
    \label{eqn:thermal_conductivity_opacity}
\end{align}
where $T$ is the temperature, $\rho$ is the density, $a =
\SI{7.5657e-16}{\joule\meter^{-3}\kelvin^{-4}}$ is the radiation
density constant, and $c$ is the speed of light.~\cite{Hayashi1962, Hansen2004}
Eq. (\ref{eqn:thermal_conductivity_opacity}) can be used to define a
conductive opacity $\kappa_\text{c}$ from the thermal
conductivity due to electrons $\lambda_\text{c}$ by analogy.
This quantity -- as calculated by Hayashi \textit{et al.} who reproduce the
work of Mestel~\cite{Mestel1950} and Lee~\cite{Lee1950} -- has been plotted in
Fig.~\ref{fig:dominant_opacities} to compare it  with radiative opacities.

As the two contributions to the full thermal conductivity are additive, the
total opacity $\kappa_\text{tot}$ is given by $1/\kappa_\text{tot} =
1/\kappa_\text{rad}+ 1/\kappa_\text{c}$, i.e., in order for $\kappa_\text{c}$
being the dominant contribution to the total opacity (see the high density
and low temperature corner of Fig.~\ref{fig:dominant_opacities}), the former
quantity has to be small compared to $\kappa_\text{rad}$.~\cite{Hansen2004}

\section{Radiography predictions}
In order to generate the transmission profiles [see
Fig.~\ref{fig:lineouts}~(b)] from extinction coefficient line-outs
[Fig.~\ref{fig:lineouts}~(c)], we assumed a perfect, spherically symmetric
implosion and a uniform and monochromatic backlighter emitting parallel x rays.
We calculated
\begin{align}
    \mathcal{T}_\nu = \exp\left(-\int \mathrm{d}x \kappa_\nu(x)
    \rho(x)\right)
    \label{eqn:Tfromkappa}
\end{align}
where the integral runs over the path of the light and might, therefore, probe
different temperature and density conditions.
To account for the limited temporal resolution of the detector, a series of
one-dimensional transmission profiles at
different times was convolved with a Gaussian gate-function
(\SI{35}{\pico\second} FWHM).
The resulting line-out has been rotated, and a spatial blur in the form of a
two-dimensional Gaussian with \SI{10}{\micro\meter} FWHM in both directions
was applied before the transmission has been multiplied with the expected
backlighter photon flux
(\SI{240}{\per\square\micro\meter\per\pico\second}
($E_\text{ph}=\SI{5.2}{\kilo\electronvolt}$) or
\SI{288}{\per\square\micro\meter\per\pico\second}
($E_\text{ph}=\SI{7.2}{\kilo\electronvolt}$) at the target's position),
corrected for attenuators shielding various
components and re-binned to the pixel-size of the detector. In a last step,
noise proportional to the individual pixels' intensity has been applied to the
data.

\bibliography{NIF_red_dwarf_methods}

\end{document}